# A new fast approach for an EEG-based Motor Imagery BCI classification

Mohammad Ali Amirabadi

Email: m_amirabadi@elec.iust.ac.ir

**Abstract**

Nowadays, Brain Computer Interface has an important role in the life quality of parallelized people. However, this technique is mainly affected by the quality of the recorded signal in each trial. This problem could be solved by rejecting low-quality trials. But developing the processing based on the recorded signal from the brain, which is a mixture of the target signal plus noise and artifact, would not be favorable in situations that all trials have low quality. This paper solves this problem by presenting a new fast algorithm for separating recorded source signals. Results indicate the improvement in classification accuracy of the proposed method compared with the classification accuracy of processing on the recorded mixture signal.

## I. Introduction

Brain Computer Interface (BCI) is a one-way information flow that translates recorded brain signals to command signals. This system mainly focuses on the way of helping paralyzed people to move without taking care of their disorders. The BCI does not use the normal output pathways of peripheral nerves and muscles [1], it only records the brain signals to prepare commands for control of external devices [2]. In motorized prostheses, movement paralysis is the main consideration. Actually, BCI provides a substitute form of communication for people with motorized impairments and helps them to send commands by using their brain activities.

Fortunately, disabled people can generate different mental states, i.e., they have the ability to perform motor imagery (MI) [3]. The movement of different parts of the body triggers different parts of the brain. In fact, subjects' ability to modulate brain activities enables BCI to detect and fulfill subjects' intentions [4], allowing paralyzed people to control a robotic arm [2]. In MI, the user is asked to make a special motion, such as the left hand. These imaginations activate some parts of the motor cortex. According to the activated area, BCI determines the imagined movement [5].

The Electroencephalography (EEG) signal often appears in response to external stimulation in the form of an electrical potential difference created by the neurons [5]. EEG is a non-invasive way to detect modulated brain signals [4], and suitable for real-time applications. The sensorimotor rhythm appears as a power change in a specific frequency range in the sensorimotor area at the moment of motion imaging. For this reason, EEG signal changes in conjunction with different movement imaginations can be used in BCI.

EEG signal is very weak because it is recorded from scalp; in addition, noise and the artifact, such as blinking or scalp movement, affect it. Therefore, features extraction from the EEG signal and accurately classify different MIs are difficult [6]. Recorded EEG signal is a mixture of many independent source signals including neuronal oscillations, event potential, spectral perturbations, and artifacts from eye movements, muscle activity, drift, and the electrode. Separating source signals could really help to solve the problem. Independent component analysis (ICA) is a well-known method for identifying independent source signals from a recorded mixture signal [7].

Common spatial patterns (CSP) is a spatial filter for oscillatory EEG components [8]. It is the most commonly used technique for feature extraction and very effective in MI classification [6]. This technique seeks to find spatial filters that maximize variance for one class and minimize variance for another class [9]. CSP usually uses all available channels (brain areas) to estimate the covariance matrix. In a specific MI task, each person has a different activated channel, and choosing the appropriate channel is a major challenge [10]. The CSP needs the covariance matrix of data. It is better to repeat a specific trial several times in order to increase accuracy. Therefore, one covariance matrix would be obtained in each repetition. EEG signal corresponding to the same task is a stationary process. Therefore, the total covariance matrix is the average of all the covariance matrices of the experiments. In EEG, averaging with equal weights over all trials does not seem to be a good idea, because different trials are contaminated by different amounts of user in-concentration, eye blink, or muscle movement artifacts.

The distraction and fatigue of subjects in the long data collection process often produce mislabeled trials. Locating the artifact data segments within a single trial is a topic of some recent works. Two spatial filters namely, CSP and ICA are widely used in MI-BCI. CSP adopts a supervised algorithm that needs plenty of labeled data to find a projection matrix for maximizing the differences between the variances of two-class EEG data. Besides, the selected data are suggested to have strong de-synchronization/synchronization. So, the CSP method requires high-quality training data with accurate labels [11]. The main disadvantage of BCI is its sensitivity to the quality of the recorded signal in each trial because, in addition to the target signal, EEG is composed of many other undesired source signals. There are many investigations considered this issue. Recently [12] tried to solve this problem; it first called trials contaminated by noise and artifact as the low-quality trials, and presented a solution for rejecting the low-quality trials. All of the processing of this work was done on the recorded mixture signal from the scalp, not on the target source signal. So, there appears another problem; what would happen if the noise and artifact sources would be dominant in all of the trials. With the previous method, in this situation, all of the trials would be considered as low quality and rejected. The solution to this problem is developing all of the processing on the target source signal. Accordingly, this paper concerns this issue and extends the previous work. Actually in this paper selection is done on the source signals which are separated by a new fast ICA algorithm. To the best of the authors' knowledge, this is the first time that a fast approximate joint diagonalization is used in ICA-based BCI application. In comparison, the obtained improvements in the experimental results over the previous work confirm the sophistication of the above arguments.

## II. Method

### A. Problem Statement and ideology of the proposed method

In this study, a novel strategy is proposed to solve the aforementioned problems. The rationale is to recognize the low-quality EEG trials. Rejecting the low-quality trials in an EEG-based MI BCI was previously investigated in [12]. However, in situations that all trials are contaminated by noise or artifact, it is difficult to distinguish trials based on the recorded EEG signal. In the following, this paper tries to address this issue. As explained earlier, the recorded signals in EEG-based MI BCI are mixtures of some different source signals. Therefore, a solution for the mentioned problem is to use the original source signals, which could be done by separating the mixture signals properly using many algorithms such as ICA. Here appears the second issue that this paper wants to address. Assuming the source signal to be $\boldsymbol{S}$, the recorded signal would be $\boldsymbol{X} = \boldsymbol{AS}$, where $\boldsymbol{A}$ is the mixture matrix. The ICA multiplies $\boldsymbol{A}^{-1}$ by $\boldsymbol{X}$ to find $\boldsymbol{S}$. But matrix inversion of high dimensional matrices has heavy processing and in low-cost applications comprises the main cost.

In this paper, both rejecting the low-quality trials by separating independent components of the recorded signal and reducing the high computation of matrix inversion in ICA are addressed. Generally speaking, diagonalization of a matrix is the best way to reduce the complexity of its inversion. Because the inverse of a diagonal matrix is just the inverse of its diagonal elements. However, the diagonalization of a matrix is not so easy. Therefore, this paper presents an approximate joint diagonalization algorithm [13]. Generally speaking, diagonalization of symmetric matrixes is easier than non-symmetric matrixes, therefore instead of $\boldsymbol{X}$, its covariance matrix is used for diagonalization. So, the proposed method would be faster than common ICA algorithms, because, in addition to using diagonalization before matrix inversion, the proposed fast diagonalization algorithm (see Appendix A) reduces the complexity of diagonalization.

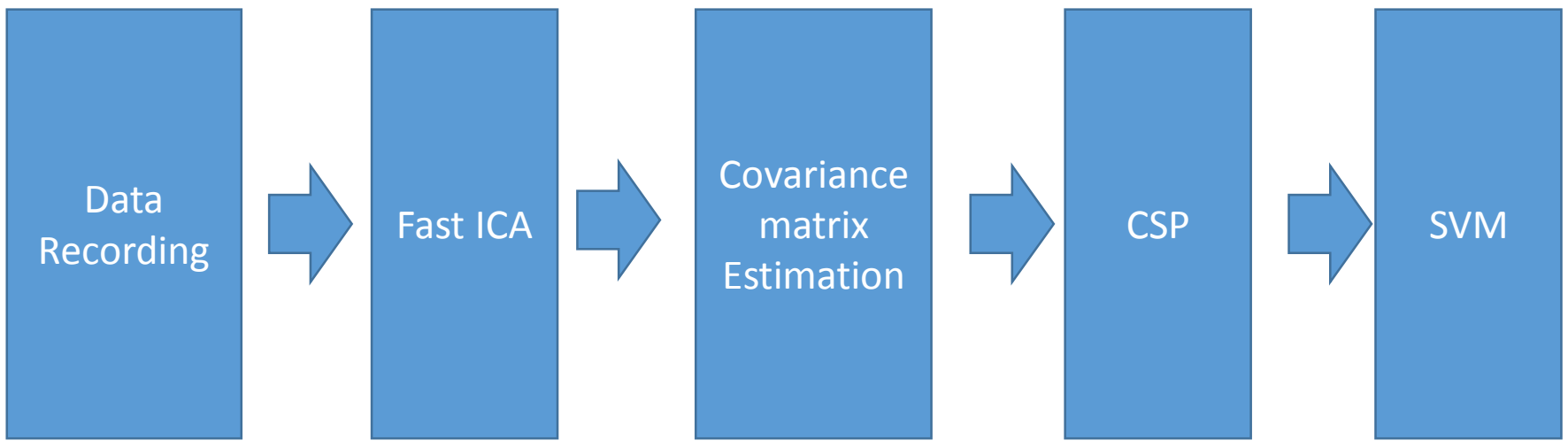

Fig. 1. Block diagram of the proposed method

The block diagram of the proposed method is shown in Fig. 1. The first block records EEG signals from the brain for each trial. The recorded signals are actually mixtures of different source signals. The second block transforms them back to the original source signals by using a fast ICA algorithm. At the third block, a fast diagonalization algorithm measures criteria for within trial qualities. According to these criteria, this block devotes sparse weights to each trial and calculates a weighted averaged covariance matrix of the recorded signal. At the forth block, the obtained covariance matrix is fed to a CSP filter. The output of this block is used as a feature for classification by Support Vector Machine in the last block. In this section, the above processing is explained in detail.

B. **Mathematical analysis of the proposed method**

Consider $\boldsymbol{X}^k = \boldsymbol{A}\boldsymbol{S}^k$ to be the recorded signal at trial $k; k = 1, \dots, K$, where $\boldsymbol{X}^k = [\boldsymbol{x}_1^k, \boldsymbol{x}_2^k, \dots, \boldsymbol{x}_N^k]$; $x_i^k \in R^M$. The goal is to estimate both $\boldsymbol{A}$ and $\boldsymbol{S}$ from $\boldsymbol{X}$. Let the within-trial covariance matrix be $\boldsymbol{C}^k$. Theoretically diagonalization process of this matrix would be $\boldsymbol{C}^k = E\left(\boldsymbol{X}^k \boldsymbol{X}^{k^T}\right) = \boldsymbol{A}E\left(\boldsymbol{S}^k \boldsymbol{S}^{k^T}\right)\boldsymbol{A}^T = \boldsymbol{A}\boldsymbol{Q}^k\boldsymbol{A}^T$. Because the source signals are independent, and the cross-correlation terms that form the off-diagonal part of $\boldsymbol{Q}^k$ are zero.

When more than two matrices are to be diagonalized, exact diagonalization may be possible if the matrices possess a certain common structure. Otherwise only approximate joint diagonalization could be used. An efficient algorithm for approximate joint diagonalization is Fast Frobenius Diagonalization [12], which is based on the second-order approximation of a cost function for the simultaneous diagonalization problem. The Fast Frobenius Diagonalization algorithm tries to find matrix $\boldsymbol{V}$ (where $\boldsymbol{V} = \boldsymbol{A}^{-1}$) that diagonalizes the given covariance matrix in the following form:

$$\boldsymbol{Q}^k = \boldsymbol{V}\boldsymbol{C}^k\boldsymbol{V}^T, \tag{1}$$

Fast Frobenius Diagonalization Algorithm iteratively finds an approximate solution for the following optimization problem:

$$\min_{\boldsymbol{V}} \sum_{k=1}^{K} \sum_{i \neq j} \left((\boldsymbol{V}\boldsymbol{C}^k\boldsymbol{V}^T)_{ij}\right)^2 \tag{2}$$

In each iteration, matrix $\boldsymbol{V}$ is updated in the following form:

$$\boldsymbol{V}_{n+1} \leftarrow (I + \boldsymbol{W}_n)\boldsymbol{V}_n \tag{3}$$

where $\boldsymbol{I}$ denotes the identity matrix, $\boldsymbol{W}_{\mathrm{n}}$ is the update matrix, constrained to have zeroes on the main diagonal, and $n$ is the iteration number.

Implementing Algorithm 1 derives the diagonalized covariance matrix ($\boldsymbol{Q} = \{\boldsymbol{Q}^1, \dots, \boldsymbol{Q}^K\}$). Therefore, the covariance matrix of source signals could be obtained by multiplying $\boldsymbol{Q}^{-1}$ by $\boldsymbol{X}$. Algorithm 2 summarizes this step (see Appendix B) [13].

In the next step, the total covariance matrix is estimated by a weighted averaging over within trial covariance matrices of the separated source signals. The assigned weights conserve the following constraints:

$$C_1: \boldsymbol{w}^T\boldsymbol{1} = 1; C_2: \boldsymbol{w}^T\boldsymbol{e}_k \geq 0, \tag{4}$$

where $\boldsymbol{w}$ is the weight vector. Actually, low-quality trials should be rejected by assigning weights obtained by solving the following $l_1$-norm optimization problem [12]:

$$\min_{\boldsymbol{w}} \frac{1}{tr(\boldsymbol{D})} \|\boldsymbol{D}\boldsymbol{w}\|_1 \tag{5}$$
$$S.t.\ \ \boldsymbol{w}^T\boldsymbol{1} = 1; \boldsymbol{w}^T\boldsymbol{e}_k \geq 0,$$

where $\boldsymbol{D} = diag[p_1, p_2, \dots, p_K]$, and $p_k$ is a scalar related to the quality of each trial. The underlying assumption behind the diagonalization is that the residue resulting from the diagonalization with respect to a low-quality trial is large. Considering $\boldsymbol{E}^k$ as the diagonalization residue, $p_k$ can take the form of $p_k = \|\boldsymbol{E}^k\|_F$, where $\boldsymbol{E}^k$ could be derived by Algorithm 1.

The ideal covariance matrix could be the equal weight average of within-trial covariance matrices. Considering this assumption, $l_1$-norm optimization changes to the following regularized $l_1$-norm optimization [12]:

$$\min_{\boldsymbol{w}} \frac{\alpha}{tr(\boldsymbol{D})} \|\boldsymbol{D}\boldsymbol{w}\|_1 + \frac{1}{2}\frac{1}{\sum_k \|\boldsymbol{Q}^k\|_F^2} \left\| \sum_k \left(\frac{1}{K} - w_k\right) \boldsymbol{Q}^k \right\|_F^2 \tag{6}$$
$$S.t.\ \ \boldsymbol{w}^T\boldsymbol{1} = 1; \boldsymbol{w}^T\boldsymbol{e}_k \geq 0,$$

where α is the regularization parameter. After some mathematical manipulation (6) becomes as

$$\min_{\boldsymbol{w}} \frac{\alpha}{tr(\boldsymbol{D})} \|\boldsymbol{D}\boldsymbol{w}\|_1 + \frac{1}{2}\frac{1}{tr(\boldsymbol{G})} \left(\boldsymbol{w} - \frac{\boldsymbol{1}}{K}\right)^T \boldsymbol{G} \left(\boldsymbol{w} - \frac{\boldsymbol{1}}{K}\right) \tag{7}$$
$$S.t.\ \ \boldsymbol{w}^T\boldsymbol{1} = 1; \boldsymbol{w}^T\boldsymbol{e}_k \geq 0,$$

where $\boldsymbol{G} = \begin{bmatrix} tr\left[\boldsymbol{Q}^1\boldsymbol{Q}^{1\boldsymbol{T}}\right] & \cdots & tr\left[\boldsymbol{Q}^1\boldsymbol{Q}^{K\boldsymbol{T}}\right] \\ \vdots & \ddots & \vdots \\ tr\left[\boldsymbol{Q}^K\boldsymbol{Q}^{1\boldsymbol{T}}\right] & \cdots & tr\left[\boldsymbol{Q}^K\boldsymbol{Q}^{K\boldsymbol{T}}\right] \end{bmatrix}$.

The projected gradient method is known as a solution for convex optimization problems, which is extensively investigated over the last decades [14]. Alternating Direction Method for Multipliers (ADMM) algorithm is used to blend the decomposability of dual ascent with the superior convergence properties of the method of multipliers. The ADMM solves the following optimization problem [15]:

$$\min_{\boldsymbol{w},\boldsymbol{v}} f(\boldsymbol{w}) + g(\boldsymbol{v}) \tag{8}$$
$$s.t.\quad \boldsymbol{A}\boldsymbol{w} + \boldsymbol{B}\boldsymbol{v} = \boldsymbol{c},$$

where $\boldsymbol{w} \in \boldsymbol{R}^n$ and $\boldsymbol{v} \in \boldsymbol{R}^m$, $\boldsymbol{A} \in \boldsymbol{R}^{p\times n}$, $\boldsymbol{B} \in \boldsymbol{R}^{p\times m}$, $\mathbf{c} \in \mathbf{R}^{\mathrm{p}}$, and $f$ and $g$ are convex functions. The Lagrangian form of the above problem is as follows:

$$L_\rho(\boldsymbol{w},\boldsymbol{v},\boldsymbol{y}) = f(\boldsymbol{w}) + g(\boldsymbol{v}) + \boldsymbol{y}^T(\boldsymbol{Aw}+\boldsymbol{Bv}-\boldsymbol{c}) + \left(\frac{\rho}{2}\right)\|\boldsymbol{Aw}+\boldsymbol{Bv}-\boldsymbol{c}\|_2^2, \quad (9)$$

where $\rho > 0$. The ADMM can solve the above problem by the following iterations:

$$\boldsymbol{w}^{n+1} = \min_{\boldsymbol{w}} L_\rho(\boldsymbol{w},\boldsymbol{v}^n,\boldsymbol{y}^n) \quad (10)$$

$$\boldsymbol{v}^{n+1} = \min_{\boldsymbol{v}} L_\rho(\boldsymbol{w}^{n+1},\boldsymbol{y}^n,\boldsymbol{v}) \quad (11)$$

$$\boldsymbol{y}^{n+1} = \boldsymbol{y}^n + \rho(\boldsymbol{Aw}^{n+1}+\boldsymbol{Bv}^{n+1}-\boldsymbol{c}) \quad (12)$$

Therefore, in order to solve (7) using ADMM, first should write its Lagrangian equivalent. The Lagrangian equivalent of (7) is in the following form:

$$\frac{\alpha}{tr(\boldsymbol{D})}\|\boldsymbol{Dw}\|_1 + \frac{1}{2}\frac{1}{tr(\boldsymbol{G})}\left(\boldsymbol{w}-\frac{\boldsymbol{1}}{K}\right)^T\boldsymbol{G}\left(\boldsymbol{w}-\frac{\boldsymbol{1}}{K}\right) + l_{C_1}(\boldsymbol{w}) + l_{C_2}(\boldsymbol{w}), \quad (13)$$

where $l_C$ is the indicator function. Furthermore, comparing (7) with (8), the following equivalencies would exit:

$$f(\boldsymbol{w}) = \frac{\alpha}{tr(\boldsymbol{D})}\|\boldsymbol{Dw}\|_1 + \frac{1}{2}\frac{1}{tr(\boldsymbol{G})}\left(\boldsymbol{w}-\frac{\boldsymbol{1}}{K}\right)^T\boldsymbol{G}\left(\boldsymbol{w}-\frac{\boldsymbol{1}}{K}\right) + l_{C_1}(\boldsymbol{w}) \quad (14)$$

$$g(\boldsymbol{v}) = l_{C_2}(\boldsymbol{v}). \quad (15)$$

Considering (10), and (13) obtains:

$$\begin{aligned}\boldsymbol{w}^{n+1} &= \min_{\boldsymbol{w}} L_\rho(\boldsymbol{w},v^n,y^n) \quad (16)\\ &= \min_{\boldsymbol{w}} \frac{\alpha}{tr(\boldsymbol{D})}\|\boldsymbol{Dw}\|_1 + \frac{1}{2}\frac{1}{tr(\boldsymbol{G})}\left(\boldsymbol{w}-\frac{\boldsymbol{1}}{K}\right)^T\boldsymbol{G}\left(\boldsymbol{w}-\frac{\boldsymbol{1}}{K}\right) + \frac{1}{2\rho}\|v^n-\boldsymbol{w}-y^n\|_2^2 + \\ &\xi(\boldsymbol{w}^T\boldsymbol{1}-1).\end{aligned}$$

Differentiating the objective function of (16) with respect to $\boldsymbol{w}$ and finding its root obtains:

$$\boldsymbol{w}^{n+1} = \left(\rho\frac{\boldsymbol{G}}{tr(\boldsymbol{G})}+\boldsymbol{I}\right)^{-1}\left[z^n-\boldsymbol{y}^n+\rho\left(\frac{\boldsymbol{G1}}{Ktr(\boldsymbol{G})}-\frac{\alpha}{tr(\boldsymbol{D})}\boldsymbol{D1}-\xi\boldsymbol{1}\right)\right]. \quad (17)$$

Substituting (17) to constraint $C_1$ in (4), and after some mathematical manipulations obtains:

$$\xi = \left(\rho \mathbf{1}^T \left(\frac{\rho \boldsymbol{G}}{tr(\boldsymbol{G})} + \boldsymbol{I}\right)^{-1} 1\right)^{-1} \left[\mathbf{1}^T \left(\frac{\rho \boldsymbol{G}}{tr(\boldsymbol{G})} + \boldsymbol{I}\right)^{-1} \left[z^n - \boldsymbol{y}^n + \rho\left(\frac{\boldsymbol{G}\mathbf{1}}{K tr(\boldsymbol{G})} - \frac{\alpha}{tr(\boldsymbol{D})}\boldsymbol{D}\mathbf{1}\right)\right] - 1\right]. \quad (18)$$

Considering (11), and (13) obtains:

$$\boldsymbol{v}^{n+1} = \min_{\boldsymbol{v}} l_{C_1}(\boldsymbol{v}) + \frac{1}{2\rho} \|\boldsymbol{v} - (\boldsymbol{w}^{n+1} + \boldsymbol{y}^n)\|_2^2 \, . \quad (19)$$

According to the definition of the indicative function, $\boldsymbol{v}$ should be in the domain of $C_2$; i.e., $\boldsymbol{v} \geq 0$. Accordingly,$\boldsymbol{w}^{n+1} + \boldsymbol{y}^n$ should be positive to minimize the objective function. In fact $\mathbf{w}^{n+1} + \mathbf{y}^n$ should be projected to the domain of $C_2$, i.e.

$$\boldsymbol{v}^{n+1} = \boldsymbol{P}_{C_2}(\boldsymbol{w}^{n+1} + \boldsymbol{y}^n). \quad (20)$$

Considering (12), can write

$$\boldsymbol{y}^{n+1} = \boldsymbol{y}^n + \boldsymbol{w}^{n+1} - \boldsymbol{v}^{n+1}. \quad (21)$$

Therefore, the target sparse weights obtain by choosing initializing points of $\mathbf{w}^0, \mathbf{v}^0, \mathbf{y}^0$ and then repeating (17), (20), (21).

## III. Experimental Results

In this section, the proposed model is conducted on an available data set. The results are then compared with the proposed model of [12] and show considerable improvements. The conducted experiment is the classification of EEG signals during MI task to show performance in accuracy with the proposed method.

The dataset 2a from BCI competition IV is used. This dataset is an EEG data recorded from 9 subjects for four different MI tasks, namely the imagination of movement of the left hand, right hand, both feet, and tongue. The data record was done for each subject at two 288 trial sessions. Subjects sat on a comfortable armchair looking at a monitor. At the beginning ($t = 0s$), a fixation cross appeared on the monitor with an acoustic warning. After two seconds ($t = 2s$), an arrow pointing either to the left, right, down or up appeared. At this time subjects performed the MI task. They should carry out the MI task until the disappearance of the fixation cross. Twenty-two electrodes were used to record EEG signals. All signals were recorded mono-polarly with reference of left mastoid and ground of right mastoid. The signals were sampled with $250\ Hz$ and bandpass-filtered between $0.5\ Hz$ and $100\ Hz$.

The following CSP methods are used for feature extraction.

1) ICA-CSP with equal weight ($w_k = 1/K$).

2) ICA-CSP with quality related weight ($w_k = \eta \|E^k\|_F^{-1}$, where $\eta$ is a normalization factor). The idea is to give a small weight to a large residue (low quality) trial.

3) ICA-CSP with sparse weight.

4) CSP with the sparse weight.

Actually, the CSP with the sparse weight is the model used by [12] and is used just for comparison. In all of these four cases, the log-variance of the output of the CSP filter is used as a feature vector. The feature is classified with Support Vector Machine (SVM).

Results are obtained by conducting 10-fold cross validation for $\alpha = 10.2, \gamma = 10^{-5}$. Results of the classification accuracy are shown in Table 1. The ICA-CSP with sparse weight has the highest classification accuracy.

The ICA-CSP with quality related weight shows a bit of accuracy improvement compared with ICA-CSP with equal weight. This shows the weight coefficients obtained by residue matrices are designed in a sophisticated case. In addition, compared with [12] (CSP with the sparse weight), the proposed model (ICA-CSP with sparse weight) shows an improvement in classification accuracy.

Table 1: Classification accuracy [%] from 10-fold cross-validation.

| Method / Subject | Proposed ICA-CSP with equal weight | Proposed ICA-CSP with quality related weight | Proposed ICA-CSP with sparse weight | CSP with the sparse weight [12] |
|---|---|---|---|---|
| 1 | 0.70503 | 0.70354 | 0.72958 | 0.73396 |
| 2 | 0.61652 | 0.63774 | 0.70036 | 0.64914 |
| 3 | 0.60446 | 0.6294 | 0.65199 | 0.6272 |
| 4 | 0.57866 | 0.58741 | 0.64223 | 0.64616 |
| 5 | 0.52057 | 0.50765 | 0.58807 | 0.58033 |
| 6 | 0.57958 | 0.56795 | 0.6736 | 0.67062 |
| 7 | 0.66616 | 0.66574 | 0.69283 | 0.69923 |
| 8 | 0.59503 | 0.6019 | 0.62658 | 0.62839 |
| 9 | 0.6503 | 0.64726 | 0.65411 | 0.65077 |
| **Avg** | **0.61292** | **0.61651** | **0.66215** | **0.65398** |

## IV. Conclusion

In classification accuracy of MI BCI could be degraded by low-quality trials. Rejecting these trials is a solution to this problem. But processing should be done on the source signals, not on the recorded signals because in the worst case scenario in which all of the trials are low quality, processing the recorded signal results in rejecting all of the trials. In this paper, a new fast approach is presented for rejecting low-quality trials. Actually, a new fast ICA algorithm is presented for separating the source signals, and then the rejection is developed on the source signal. In comparison with the previous work, results indicate the improvement of classification accuracy of the proposed method.

## Appendix A

Algorithm 1, Fast Approximate Joint Diagonalization

Input $\boldsymbol{C} = \{\boldsymbol{C^1}, \dots, \boldsymbol{C^K}\}, ;$

$\boldsymbol{W_1} = \boldsymbol{0}, \boldsymbol{V_1} = \boldsymbol{I};$

```
       while error > epsilon & n < 1000

% Compute W

For k=1:K
     z(i,j) = z(i,j)+C(i,i,k)*C(j,j,k);
     y(i,j) = y(i,j)+0.5*C(j,j,k)*(C(i,j,k)+conj(C(j,i,k)));

end

    W(i,j) = (z(j,i)*y(j,i)-z(i,i)*y(i,j))/(z(j,j)*z(i,i)-z(i,j)^2);
    W(j,i) = ((z(i,j)*y(i,j)-z(j,j)*y(j,i))/(z(j,j)*z(i,i)-z(i,j)^2));
    [f,e] = log2(norm(W,'inf'));
    s = max(0,e-1);
    W = W/(2^s );

% Compute update

   V = (I+W)*V;
   V=diag(1./sqrt(diag(V*V')))*V;

   C= A*C*V';

for k=1:K
    f = f + trace((V*C(:,:,k)*V')'* V*C(:,:,k)*V') -
trace(V*C(:,:,k)*V'.*V*C(:,:,k)*V');
end

% convergence

   error = abs(f(n)-f(n-1));
     n=n+1;
end
```

## Appendix B

Algorithm 2, ICA

```
Input  X
% tau -- array of time-delays, default: tau=[ 0 1];

X = X - mean(x);
C= zeros(N, N, 2);

for t=1:2
  C0= x(:,1:T-tau(t))*x(:,1+tau(t):T)' / (T-tau(t)-1);
  C(:,:,t)= (C0+C0')/2;
end

  Q = Algorithm 1(C)
```